\documentclass[aps,floats,floatfix,preprint,prl]{revtex4}

\usepackage[dvips]{graphicx}
\usepackage[usenames,dvipsnames]{color} 

\begin{document}

\title{Solvation vs. freezing in a heteropolymer globule}
\author{Phillip L.~Geissler}
\affiliation{Department of Chemistry, Massachusetts Institute of
Technology, Cambridge, MA 02139} 
\author{Eugene I.~Shakhnovich}
\affiliation{Department of Chemistry and Chemical Biology, Harvard
University, Cambridge, MA 02138}
\author{Alexander Yu.~Grosberg}
\affiliation{Department of Physics, University of Minnesota,
Minneapolis, Minnesota 55455}

\begin{abstract}
We address the response of a random heteropolymer to preferential
solvation of certain monomer types at the globule-solvent interface.
For each set of monomers that can comprise the molecule's surface, we
represent the ensemble of allowed configurations by a Gaussian
distribution of energy levels, whose mean and variance depend on the
set's composition.  Within such a random energy model, mean surface
composition is proportional to solvation strength under most
conditions.  The breadth of this linear response regime arises from
approximate statistical independence of surface and volume energies.
For a diverse set of monomer types, the excess of solvophilic monomers
at the surface is large only for very strong solvent preference, even
in the ground state.
\end{abstract}

\maketitle 

A polymer chain collapses into a compact globular state in poor
solvent.  A chain with quenched sequence of chemically different units
can further undergo a freezing transition, in which the freedom of
chain shape fluctuations is sacrificed for the choice of optimal
spatial contacts between monomers.  This freezing, or folding, is
subject to constraints imposed by chain connectivity, quenched
sequence, and excluded volume.  The effects of frustration due to
these constraints are well understood
\cite{shakh_gutin,freezing_shakh,freezing_gros}.   However well developed,
current theories of heteropolymer freezing ignore one obvious fact,
namely that some chain segments are more favorably solvated than
others.  By contrast, much of the protein literature presumes
preferential solvation to be a leading determinant of tertiary
structure.  It is commonly held that a protein's surface is composed
of hydrophilic units, while hydrophobic units are invariably buried in
the core.

For heteropolymers in general, it is clear that energy gained through
preferential exposure of solvophilic units comes at a cost.
Constraining particular units to the globule surface restricts the
selection of contacting monomer pairs inside the globule, exacerbating
frustration.  In other words, when the sequence of units has not been
designed in an intelligent way, as is the case for the random sequence
heteropolymer, preferential exposure may significantly reduce the
availability of low energy conformations.  The question thus arises:
what is the effect of solvation on heteropolymer freezing, or more
specifically, how large an excess of solvophilic units at the globule
surface is consistent with freezing?

This question was first discussed by two of us \cite{GS} in the context of
studies of mechanical stretching of heteroplymers.  
Using a replica approach, we found that
solvent preference of strength $\Gamma$ for particular monomers at the
surface lowers ground state energy $E_{\rm g}$ by an amount $\sim K
\Gamma^2 /T_{\rm fr}$. Here, $K \sim N^{2/3}$ is the number of
monomers exposed to the solvent, $N$ is the number of monomers
comprising the molecule, and $T_{\rm fr}$ is the freezing temperature
below which the ground state dominates.  For strong solvation this
approach apparently fails.  In particular, $K \Gamma^2 / T_{\rm fr}$
can exceed the maximum possible solvation energy (without distortion
of globule shape), $K \Gamma$, corresponding to a completely
solvophilic surface.

This Letter describes a more comprehensive treatment of solvation
based on the Random Energy Model (REM) of Derrida
\cite{derrida}. It is well known 
\cite{shakh_gutin,freezing_shakh,freezing_gros,bryng_wolynes} that this
simplest model of freezing in spin glasses captures remarkably
well the essential features of heteropolymer freezing in three
dimensions. The mapping of heteropolymer problem on the REM is
achieved by approximating energies of all $M=e^{s N}$ different
conformations of a random sequence heteropolymer as $M$
independent random variables drawn from the Gaussian distribution
$w(E) \propto \exp \left[ - \left( E - \overline{E} \right)^2 / 2
N \Delta^2 \right]$.  In the volume approximation, when energy of
every conformation is solely due to $\sim N$ monomer-monomer
contacts, this distribution is fully determined by the mean
$\overline{B}$ and variance $\delta \! B^2$ of contact energies,
so that $\overline{E} = N \overline{B}$ and $\Delta^2 = \delta \!
B^2$.

The simplest way to incorporate surface into this picture is to
imagine that contacts between surface monomers and solvent are also,
in effect, statistically independent random variables.  The variance
of surface energy, $K \Gamma^2$, then adds to that of volume energy.
We use the saddle point of the partition function, $Z = e^{sN} \int \!
dE \, w(E) e^{-E/T}$, to estimate the energy of representative
conformations, $E = \overline{E} - N \delta\! B^2 / T - K \Gamma^2 /
T$.  The lower bound of the spectrum is reached when $e^{sN} w(E)
\simeq 1$, yielding a typical ground state energy:
\begin{equation}
E_{\rm g}^{(\rm typ)} \simeq \overline{E} - \sqrt{2s} N \Delta \simeq
\overline{E} - \sqrt{2s} N \delta\! B - \sqrt{s \over 2} {\Gamma^2
\over \delta\! B}.
\label{equ:ground_ind}
\end{equation}
Correspondingly, $T_{\rm fr} \simeq \delta\! B / \sqrt{2s}$.  The
final term in Eq.~\ref{equ:ground_ind}, i.e., the change in ground
state energy due to solvation, is (within a factor of order unity) $-K
\Gamma^2 / T_{\rm fr}$, just as found in the replica approach of
Ref.~\cite{GS}.

There are several reasons to be skeptical of the suggested
independence of surface and volume energies.  First, removing a
specific set of monomers from the globule interior to the surface
modifies the distribution of contacting monomers.  Secondly, there are
a finite number of solvophilic monomers in a given molecule (possibly
fewer than $K$).  When a large fraction is placed on the surface, the
supply of solvophilic monomers is strongly depleted, and solvation
energy saturates.  Finally, certain choices of surface monomers
constrain configuration space more strongly than others.

We examine these effects using a model in which each monomer is
labeled by a quenched variable $\sigma$.  When a monomer with label
$\sigma$ resides on the surface, it is assigned a solvation energy
$-\Gamma \sigma$.  Solvophilic species are thus characterized by
$\sigma > 0$, while $\sigma < 0$ for solvophobic species.  In its
total effect the solvent preference $\Gamma$ can be viewed as an
external field that couples linearly to net surface composition,
$C_{\rm surf} \equiv \sum_{i\in {\rm surf}}\sigma_i$.  Within the
globule, a contacting pair of monomers, of type $\sigma$ and
$\sigma'$, is ascribed energy $B_{\sigma \sigma'} = \overline{B} +
\delta\!B \, \sigma \sigma'$.  For simplicity we restrict attention to
distributions of monomer types, $p(\sigma)$, with zero mean and unit
variance:
\begin{equation}
\int d\sigma p(\sigma) \sigma = 0, \qquad \int d\sigma p(\sigma)
\sigma^2 = 1.
\label{equ:p}
\end{equation}

Imagine that a certain set of monomers is constrained to sit on the
surface.  We denote this set as $G$.  In our model energetic
consequences of such a constraint depend only on the distribution,
$f(\sigma)$, of monomer types in $G$.  For example, the effective
distribution of contacting monomers (i.e., those remaining inside the
globule when $G$ is removed), $p_{\rm eff}(\sigma)$, may be written as
$p_{\rm eff}(\sigma) = p(\sigma) + {K \over N} [p(\sigma) -
f(\sigma)]$.  The effective mean and variance of contact energies are
then $\overline{B}_{\rm eff} = \overline{B} + (K/N) \alpha_G$ and
$\delta\!  B_{\rm eff} = \delta\! B + (K/N) \beta_G$, respectively.
For distributions satisfying Eq.~\ref{equ:p}, $\alpha_G=0$ and
$\beta_G = 2 \delta\! B^2 [1- \int d\sigma \sigma^2 f(\sigma) ]$.
Similarly, the solvation energy per surface monomer is $\gamma_G =
-\Gamma \int d\sigma \sigma f(\sigma)$.

We express the number of accessible conformations when all monomers in
$G$ are confined to the surface as $M_G \equiv e^{sN - K \omega_G}$.
Here, $\omega_G$ is the entropy loss per surface monomer for
particular of $G$.  Though smaller than $M$, $M_G$ is still
exponentially large in $N$.  In general $\omega_G$ is not simply a
functional of $f(\sigma)$, but is instead a complicated function of
$G$.  We will assume that for any specific $f(\sigma)$, the average of
$\omega_G$ over all consistent realizations of $G$ is a constant
independent of $f(\sigma)$.  In order to recover the appropriate total
number of conformations after summing over $G$, we choose this
constant to be $\overline{\omega} = K^{-1}\ln{N \choose K} \simeq
\ln{(N e/K)}$.

We consider a separate REM for each possible choice of surface $G$.
In doing so, we assume that allowed conformations in the corresponding
subensembles are sufficiently diverse that their energies are Gaussian
distributed, with
\begin{equation}
w_G(E) \propto \exp{\left[-{\left(
E - N\overline{B} - K(\alpha_G + \gamma_G) \right)^2 \over 
2N\delta\! B^2 + 2K\beta_G}
\right]}.
\label{equ:REM_G}
\end{equation}
Ultimately, we must reconstruct the full ensemble of compact chain
fluctuations by superposing all possible subensembles, i.e., by
summing over $G$.  This convolution of REMs, each representing a
distinct choice of $G$, constitutes our caricature of a random
heteropolymer with solvated surface.

Consider the ground state of the full ensemble, i.e., the lowest
of subensemble ground state energies, $E_{\rm g} = \min_G{E_{\rm
g}(G)}$.  Interfacial energy clearly favors a solvophilic surface, but
does it yield the lowest ground state?  Let us first examine a {\em
typical} value of $E_{\rm g}(G)$ for specific $G$.  The
condition $M_G w_G[E_{\rm g}^{(\rm typ)}(G)] \simeq 1$ yields:
\begin{eqnarray}
E_{\rm g}^{(\rm typ)}(G) &\simeq& N\overline{B} - N\sqrt{2s}\delta\! B +
K \epsilon_{\rm surf}(G), \\ \epsilon_{\rm surf}(G) &=& \alpha_G +
\gamma_G + {\delta\! B \over \sqrt{2s}}\overline{\omega} -
\sqrt{s\over 2} {\beta_G \over \delta\! B}.
\end{eqnarray}
This most probable ground state energy is indeed minimized by an
exclusively solvophilic choice of $G$.  There are many distinct
choices of $G$, however, leading to the same value of $E_{\rm g}^{(\rm
typ)}(G)$.  It is therefore crucial to account for variations in
$E_{\rm g}(G)$ among similar subensembles.  According to the
statistics of extreme values\cite{bouchaud}, the probability that the
lowest energy in a particular subensemble deviates from $E_{\rm
g}^{(\rm typ)}(G)$ by an amount $\delta\!E_{\rm g}(G) = E_{\rm g}(G)-
E_g^{(\rm typ)}(G)$ is
\begin{equation}
{\cal W}[\delta\! E_{\rm g}(G)] = \exp{\left[{\delta\!E_{\rm
g}(G) \over T_{\rm fr}} - \exp{\left({\delta\! E_{\rm g}(G) \over T_{\rm
fr}}\right)} \right]}.
\end{equation}
Compared to the Gaussian distribution of energies within a
subensemble, ${\cal W}[\delta\!E_{\rm g}(G)]$ decays very slowly for
$\delta\!E_{\rm g}(G) < 0$.  When many subensembles share a common
value of $E_{\rm g}^{(\rm typ)}(G)$, their range of $E_{\rm g}(G)$
will be broad.

The vast majority of subensembles have unremarkable surface energy.
The number with $|\epsilon_{\rm surf}(G)| \ll K$ and $\delta\!E_{\rm
g}(G)={\cal E}$ is thus roughly $e^{\overline{\omega} K} {\cal
W}({\cal E})$.  Since the tail of ${\cal W}[\delta\! E_{\rm g}(G)]$ is
exponential, we expect $O(1)$ of the subensembles with insignificant
surface energy to have $|\delta\!E_{\rm g}(G)| = O(K)$.  In other
words, the variations in volume energy among these subensembles are
comparable in magnitude to the largest possible surface energy.  This
result may be viewed as the consequence of an effective entropy that
remains important even at low temperature.  The collection of
subensembles with appreciable surface energy is much smaller than
$e^{\overline{\omega}K}$, and its entropy is correspondingly low.  The
ground state surface is uniformly solvophilic only when solvent
preference is strong enough to offset this entropic cost.

Because $w_G(E)$ depends only on $f(\sigma)$, it is natural to group
all subensembles with the same number density of monomer types.  We
have shown that accounting for the disparity in sizes of these groups
is essential.  The number of ways to choose $K$ monomers with
distribution $f(\sigma)$ from a pool of $N$ monomers with distribution
$p(\sigma)$ is $e^{N s\{f \}}$, where
\begin{equation}
s\{f\} = - \int d \sigma p(\sigma) [\phi \ln{\phi} +
(1-\phi)\ln{(1-\phi)}].
\label{equ:grouping_entropy}
\end{equation}
The density $\phi(\sigma) \equiv K f(\sigma) / N p(\sigma)$ and its
corresponding entropy, $s\{f\}$, are precisely those relevant for
Langmuir adsorption of an ideal gas mixture onto $K$ distinguishable
sites.

At and above the freezing temperature, equilibrium of a 
subensemble group is dominated by the saddle point of the partition
function
\begin{equation}
Z\{f\} = e^{Ns - K\overline{\omega} + N s\{f \}} \int dE\, w_G(E)
e^{-E/T}.
\end{equation}
The group free energy, $F\{f\} = -T\ln{Z\{f\}}$, is then
\begin{eqnarray}
F\{f\} \simeq \overline{F} + K T \left[ -{\delta \! B^2 \over T^2} +
\overline{\omega} \right] + N T \int d\sigma p(\sigma) && \nonumber \\
&&
\hspace{-6.5cm} \times \left\{ \phi \left[ \eta(\sigma) -
\ln{\left({1-\phi \over \phi } \right)} \right] + \ln{(1-\phi)}
\right\},
\label{equ:group}
\end{eqnarray}
where 
\begin{equation}
\eta(\sigma) = {\delta \! B^2 \over T^2} \sigma^2 - 
{\Gamma \over T} \sigma .
\label{equ:eta}
\end{equation}
Volume terms independent of $f(\sigma)$ have been collected as
$\overline{F}/N = \overline{B} -Ts - \delta \! B^2 / 2 T$.  According
to Eqs.~\ref{equ:group} and \ref{equ:eta}, the binding energy in our
analogy to Langmuir adsorption varies with particle type $\sigma$ as
$(\delta\! B^2 / T) \sigma^2 - \Gamma \sigma$.

The full partition function of the polymer, a sum over all $Z\{f\}$,
is dominated by the subensemble group with lowest free energy:
\begin{equation}
Z = \sum_{f(\sigma)} Z\{f\} \simeq Z\{f^*\}.
\end{equation}
We calculate the optimal surface distribution, $f^*(\sigma)$,
variationally, using a Lagrange multiplier to enforce
proper normalization of $\phi(\sigma)$.  We thereby obtain
\begin{equation}
f^*(\sigma) = { {N \over K} p(\sigma) \over 1 + \Lambda
e^{\eta(\sigma)} },
\label{equ:opt_dist}
\end{equation}
where the constant $\Lambda$ is determined by normalization
\begin{equation} 
\int d\sigma { {N \over K} p(\sigma) \over 1 + \Lambda
e^{\eta(\sigma)} } = 1.
\label{equ:fugacity}
\end{equation}
Finally, evaluating $F\{f\}$ at $f^*(\sigma)$ yields our approximation
for the total free energy $F = \overline{F} + F_{\rm surf}$, with
\begin{eqnarray}
{F_{\rm surf} \over K} &\simeq& T \left[ {\delta \! B^2 \over T^2} - 1
+ \ln{\left(\Lambda \int d\sigma {p(\sigma) \over 1 + \Lambda
e^{\eta(\sigma)} } \right)} \right.  \nonumber \\ && \qquad \qquad -
\left.  { \int d\sigma p(\sigma) \ln{\left( {\Lambda e^{\eta(\sigma)}
\over 1 + \Lambda e^{\eta(\sigma)}}\right)} \over \int d\sigma
{p(\sigma) \over 1 + \Lambda e^{\eta(\sigma)} } } \right].
\label{equ:saddle}
\end{eqnarray}
Eqs.~\ref{equ:opt_dist}--\ref{equ:saddle}, appropriate for $T \geq
T_{\rm fr}$, are our principal results.  They express the equilibrium
distribution of monomer types on the polymer surface and the
corresponding interfacial free energy density in terms of model
parameters and an effective fugacity for surface monomers, $\Lambda$.

In order to make these results concrete we consider some limiting
cases and specific forms of $p(\sigma)$.  First, let us assume that
preferential solvation does not lead to a significant depletion of any
monomer type inside the globule, so that $K f(\sigma) \ll N p(\sigma)$
for every $\sigma$.  Then, Eq.~\ref{equ:opt_dist} requires that
$\Lambda e^{\eta(\sigma)} \gg 1$, simplifying the above expressions to
yield $f(\sigma) \propto p(\sigma) \exp{[-\eta(\sigma)]}$ and
\begin{equation}
{F_{\rm surf} \over K} \simeq 
-T \ln{\left[\int d\sigma p(\sigma) \exp{\left({\delta\!B^2 \over
T^2}-\eta(\sigma) \right)} \right]}.
\label{equ:weak}
\end{equation}

To simplify this result even further, let us consider a binary
distribution, $p(\sigma) = (1/2)[\delta(\sigma+1) +
\delta(\sigma-1)]$, which corresponds to the minimum of chemical
diversity.  In this case depletion is invariably weak, since taking
$K$ monomers away to the surface cannot exhaust the total stock,
$N/2$, of either monomer type.  Eq.~\ref{equ:weak} then trivially
yields
\begin{equation}
{F_{\rm surf} \over K} = -T\ln{\left[\cosh{\left( {\Gamma \over T}
\right)} \right]}.
\label{equ:surf_binary}
\end{equation}
For $\Gamma/T \ll 1$, $F_{\rm surf} \simeq -K\Gamma^2 / 2T$, precisely
as obtained by assuming statistical independence of surface and
volume.  Since net surface composition is conjugate to solvation
strength, its equilibrium value may be computed by differentiating
Eq.~\ref{equ:surf_binary} with respect to $\Gamma$, yielding:
\begin{equation}
\langle C_{\rm surf} \rangle_{\Gamma} = K \tanh{\left({\Gamma \over T}
\right)} \simeq {K \Gamma \over T}.
\label{equ:linear_response}
\end{equation}
In this limit net surface composition is proportional to the ``field''
$\Gamma$.  The above results may therefore be understood in simple
terms as a manifestation of linear response.  From
Eq.~\ref{equ:linear_response} we identify a susceptibility $\chi
\simeq K/T$, corresponding to surface fluctuations of size $\langle
C_{\rm surf}^2 \rangle_{\Gamma=0} = K$ in the absence of solvation.
In other words, the excess of solvophilic monomers at the surface is
governed by $K$ effectively independent random variables.  This simple
behavior results directly from the prevalence of variations in volume
energy over surface interactions.  But when $\Gamma \gtrsim T$,
solvation wins out.  Linear response then breaks down due to
saturation, as $F_{\rm surf}/K$ and $C_{\rm surf}$ approach their
limiting values of $-\Gamma$ and $K$.

Properties of the ground state are obtained by evaluating
Eq.~\ref{equ:linear_response} and \ref{equ:surf_binary} at $T = T_{\rm
fr}$.  Dependence on the interaction parameter $\delta\!B$ is implicit
(through $T_{\rm fr}$) below the freezing transition.  For $T > T_{\rm
fr}$, however, surface response is insensitive to $\delta\!B$.  In
particular, $\beta_G$ vanishes, since no binary choice of $f(\sigma)$
can change the second moment of the contact energy distribution.  As a
consequence, surface and volume behave independently for arbitrary
$\Gamma$.

The opposite extreme of monomer diversity is described by a smooth
form of $p(\sigma)$, describing a continuous variety of chemical
identities.  We take a Gaussian distribution, $p(\sigma) \propto
\exp{(-\sigma^2/2)}$ as a simple example.  For weak solvation, $\Gamma
/ T \lesssim 1$, $p(\sigma)$ is nowhere significantly depleted, and
Eq.~\ref{equ:weak} remains an appropriate approximation.  Gaussian
integration yields
\begin{equation}
{F_{\rm surf} \over K} = -{\Gamma^2 \over 2T} - {\delta \! B^4 \over
T^3}
\label{equ:gaussian}
\end{equation}
to leading order in $\delta\!B / T$.  (The basic assumption that
monomer contacts are statistically independent is plausible only for
$\delta\!B / T_{\rm fr} = \sqrt{2s} \ll 1$\cite{REM_applic}.)  The
first term in Eq.~\ref{equ:gaussian} again reflects linear response.
The second term describes the benefit in monomer contact energy due to
partial removal of some monomer types from the globule interior.  This
effect is independent of solvation strength to leading order and
dominates interfacial free energy for very small $\Gamma$.

For such a diverse set of monomer types, surface response saturates
only when a molecule's supply of the most solvophilic type is
exhausted.  Assuming weak depletion is clearly inappropriate here.  A
maximally solvophilic surface is obtained when $K f(\sigma) = N
p(\sigma)$ for $\sigma \geq \sigma_{\rm max}$, and $f(\sigma)=0$ for
$\sigma < \sigma_{\rm max}$.  The cutoff point $\sigma_{\rm max}$ is
determined by normalization:
\begin{equation}
\int^{\infty}_{\sigma_{\rm max}} d\sigma p(\sigma) = {K \over N}.
\label{equ:max}
\end{equation}
For Gaussian $p(\sigma)$, Eq.~\ref{equ:max} gives $\sigma_{\rm max} =
\sqrt{2 \ln{(N/K)}} \simeq \sqrt{(2/3) \ln{N}}$.  Because this choice
of surface composition uniquely specifies a monomer set $G$, the
associated entropy $s\{f\}$ vanishes.  Free energy is then easily
estimated from Eq.~\ref{equ:group}, giving $F_{\rm surf}/K \simeq -
\Gamma \sigma_{\rm max}$.  Comparing this result with the free energy
of linear response, we estimate that saturation occurs around $\Gamma
\simeq 2 T \sigma_{\rm max} \sim T \sqrt{\ln{N}}$.  Reaching this
crossover may thus require much stronger solvation, and result in more
favorable surface energy, than in the binary case.  The relevant
distinction between these distributions is the existence of extremely
solvophilic monomers, whose small numbers entail considerable entropic
cost in constraining them to the surface.

\begin{figure}
\includegraphics[width=8cm]{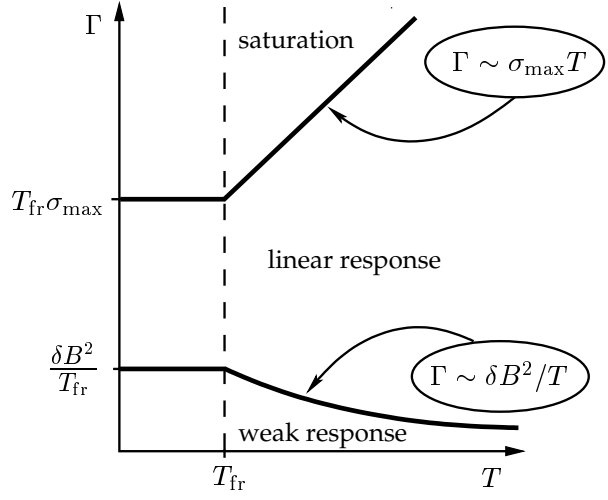}
\caption{Response of a random heteropolymer to surface solvation,
shown in the plane of temperature $T$ and solvation strength $\Gamma$.
Crossover lines are the result of equating free energies, or
ground state energies for $T<T_{\rm fr}$.  For a
binary distribution of monomer types, the weak response regime is
absent, and $\sigma_{\rm max}=1$.  
}
\label{fig:diagram}
\end{figure}

Fig.\ref{fig:diagram} summarizes the mechanisms of surface response we
have identified.  These results support a view of surface solvation
energy and volume energy as statistically independent random
variables.  In particular, the linear response corresponding to this
notion is valid over a wide range of temperature and solvation
strength.  Saturation at large $\Gamma$, though a nonlinear effect,
does not truly arise from correlation of surface and volume.  It is
instead a consequence of the finitude of surface area or of the number
of solvophilic monomers.  The regime of weak response, in which
$F_{\rm surf}/K \sim \delta\!B^4 / T^3$, does reflect coupling of
surface and volume.  But it involves monomer contact energies alone,
as indicated by insensitivity to $\Gamma$.  Within our model,
contributions from more intimate connections between surface and
volume are small compared to unity when $K \ll N$.

The diversity of amino acid monomers comprising proteins lies
somewhere between those of binary and Gaussian distributions.  The
surface behavior we have described should thus be relevant for chains
of these units arranged in random sequence.  Specifically, we predict
that preferential solvation must be much larger than typical thermal
fluctuations in order to stabilize a strictly solvophilic surface.
Sequences found in nature, however, are not random in at least one
respect important to freezing.  Their ground states lie well below the
effective continuum of non-native energies.  The influence of this
energy gap on surface solvation requires a consideration of sequence
design that is beyond this discussion.

P.L.G. is an M.I.T. Science Fellow.

\bibliographystyle{prsty}

\end{document}